\documentclass{ws-procs9x6}

\begin{document}

\title{Exclusive pion electroproduction and 
transversity}

\author{P. Kroll}

\address{Fachbereich Physik, Universit\"at Wuppertal,\\
Wuppertal, D-42097, Germany\\
$^*$E-mail: kroll@physik.uni-wuppertal.de}

\begin{abstract}
In this talk it is reported on an analysis of hard exclusive  $\pi^+$
electroproduction within the handbag approach. Particular emphasis is
laid on single-spin asymmetries. It is argued that a recent HERMES measurement
of asymmetries measured with a transversely polarized target clearly indicate 
the occurrence of strong contributions from transversely polarized photons. 
Within the handbag approach such $\gamma^{\,*}_T\to \pi$ transitions are
described by the transversity GPDs accompanied by a twist-3 pion wave
function. It is shown that this approach leads to results on cross sections
and single-spin asymmetries in fair agreement with experiment.
\end{abstract}


\bodymatter

\section{Introduction}
In this article it will be reported upon an analysis of hard exclusive
electroproduction of positively charged pions~\cite{GK5} within the 
frame work of the so-called handbag approach which offers a partonic 
description of meson electroproduction provided the virtuality of the 
exchanged photon, $Q^2$, is sufficiently large. The theoretical basis 
of the handbag approach is the factorization of the process amplitudes 
in hard partonic subprocesses and soft hadronic matrix elements,  
parameterized as generalized parton distributions (GPDs), as well as wave 
functions for the produced mesons, see Fig.\ \ref{fig:1}. In collinear 
approximation factorization has been shown~\cite{rad96,col96} to hold 
rigorously for exclusive meson electroproduction in the limit
$Q^2\to\infty$. It has also been shown that the transitions from a
longitudinally polarized photon to the pion, $\gamma^{\,*}_L\to \pi$,
dominates at large $Q^2$. Transitions from transversely polarized photons 
to pions, $\gamma^{\,*}_T\to \pi$, are suppressed by inverse powers of the
hard scale.
 
However, as has been argued in Ref.\ \refcite{GK5}, $\gamma^{\,*}_T\to \pi$
transitions are quite large at experimentally accessible values of $Q^2$ 
which are typically of the order of a few ${\rm GeV}^2$. This follows from 
data of asymmetries measured with a transversely polarized 
target~\cite{Hristova} and is also seen in the transverse cross section 
measured by the $F_\pi-2$ collaboration~\cite{horn06}. It is demonstrated 
in Ref.\ \refcite{GK5} that within the handbag approach, these  
$\gamma^{\,*}_T\to \pi$ transitions can be calculated as a twist-3 effect 
consisting of the leading-twist helicity-flip GPDs~\cite{diehl01,hoodbhoy} 
combined with the twist-3 pion distribution amplitude~\cite{braun90}. In 
the following the main ideas of the approach advocated for in Ref.\ 
\refcite{GK5} will be briefly described and some of the results will be 
discussed and compared to experiment.

\section{The handbag approach}   
\begin{figure}[b]
\psfig{file=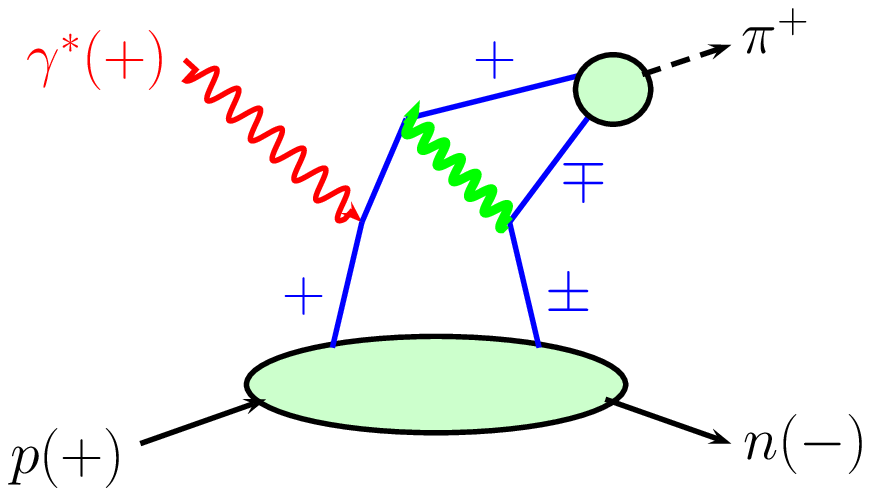,bb=105 513 358 658,width=0.52\textwidth,clip=true}
\psfig{file=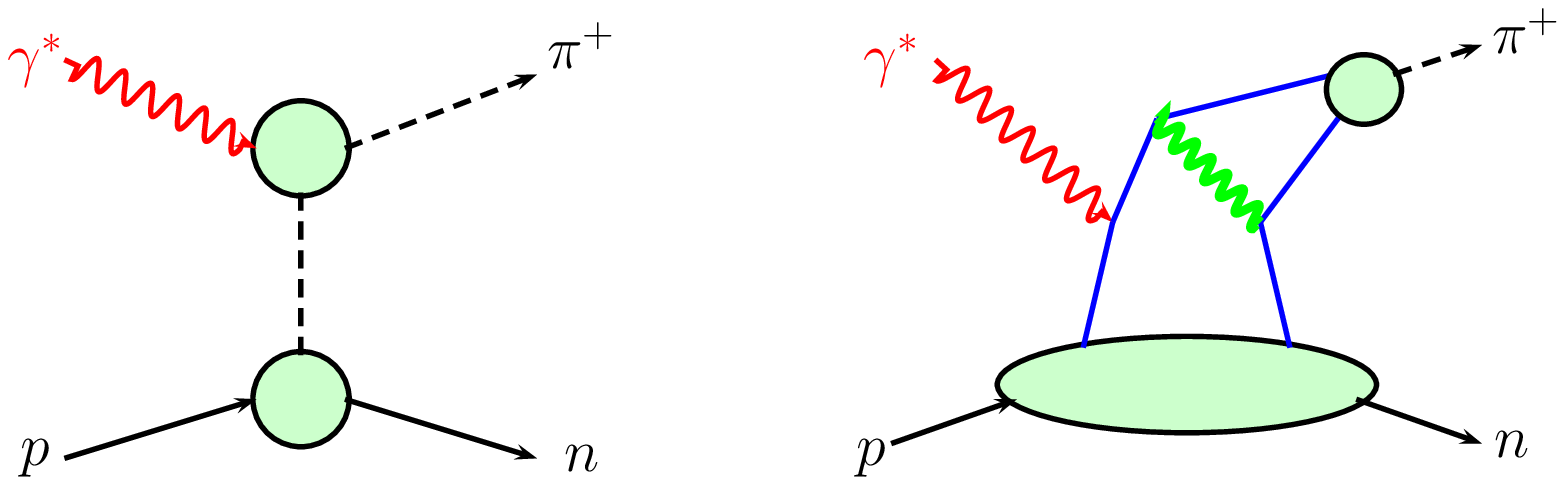,bb=147 519 325 652,width=0.40\textwidth,clip=true}
\caption{\label{fig:1} The pion exchange graph and a typical lowest order Feynman graph 
for pion electroproduction. The signs indicate helicity labels for the
contribution from transversity GPDs to the amplitude ${\cal M}_{0-,++}$, see text.}
\end{figure}

Within the handbag approach the amplitudes for pion electroproduction through
longitudinally polarized photons read
\begin{eqnarray}
{\cal M}^{\pi^+}_{0+,0+} &=& \sqrt{1-\xi^2}\, \frac{e_0}{Q}
                             \,\Big[\langle \widetilde{H}^{(3)}\rangle
-\frac{\xi^2}{1-\xi^2}\langle \widetilde{E}^{(3)}\rangle 
  - \frac{2\xi mQ^2}{1-\xi^2}\frac{\rho_\pi}{t-m_\pi^2}\Big]\,,\nonumber\\
{\cal M}^{\pi^+}_{0-,0+} &=& \frac{e_0}{Q}\,\frac{\sqrt{-t^\prime}}{2m}\,\Big[ \xi 
\langle \widetilde{E}^{(3)}\rangle + 2mQ^2\frac{\rho_\pi}{t-m_\pi^2}\Big]\,.
\label{eq:L-amplitudes}
\end{eqnarray}
Here, the usual abbreviation $t^\prime=t-t_0$ is employed where 
$t_0=-4m^2\xi^2/(1-\xi^2)$ is the minimal value of $t$ corresponding to
forward scattering. The mass of the nucleon is denoted by $m$ and the skewness
parameter, $\xi$, is related to Bjorken-$x$ by 
\begin{equation}
\xi =\frac{x_{\rm Bj}}{2-x_{Bj}}\,.
\end{equation}
Helicity flips at the baryon vertex are taken into account since they are only
suppressed by $\sqrt{-t^\prime}/m$. In contrast to this, effects of order 
$\sqrt{-t^\prime}/Q$ are neglected. The last term in each of the above
amplitudes is the contribution from the pion pole (see Fig.\ \ref{fig:1}). Its 
residue reads
\begin{equation}
\rho_\pi = \sqrt{2}g_{\pi NN} F_\pi(Q^2) F_{\pi NN}(t^\prime)\,,
\label{residue}
\end{equation}
where $g_{\pi NN}$ is the familiar pion-nucleon coupling constant. The
structure of the pion and the nucleon is taken into account by form factors,
the electromagnetic one for the pion, $F_\pi(Q^2)$, whereby the small
virtuality of the exchanged pion is as usual ignored, and $F_{\pi NN}(t)$ for 
the $\pi$-nucleon vertex. The pion-pole term has been used in essentially this
form in the measurement of the pion form factor~\cite{horn06}. In contrast to
other work on hard exclusive pion electroproduction ( an exception is Ref.\
\refcite{bechler}) the full pion form factor is taken into account this way 
and not only its so-called perturbative contribution which only amounts to
about a third of its experimental value. It is to be stressed that the pion
pole also contributes to the amplitudes for transversely polarized photons. 
However, these contributions are very small~\cite{GK5}.

The convolutions $\langle F\rangle$ in (\ref{eq:L-amplitudes}) have been worked
out in Ref.\ \refcite{GK5} with subprocess amplitudes calculated within the
modified perturbative approach~\cite{sterman}. In this approach the quark
transverse momenta are retained in the subprocess and Sudakov suppressions are
taken into account. The partons are still emitted and re-absorbed by the nucleon 
collinearly, i.e.\ we still have collinear factorization in GPDs and hard
subprocess amplitudes. It has been shown~\cite{GK1} that within this variant
of the handbag approach the data on cross sections and spin density matrix 
elements for vector-meson production are well fitted for small values of skewness
($\;\xi\simeq x_{Bj}/2\,{\raisebox{-4pt}{$\,\stackrel{\textstyle
                                                         <}{\sim}\,$}}\, 0.1\;$).

\section{$\gamma^{\,*}_T\to \pi$ transitions}
\begin{figure}[ht]
\psfig{file=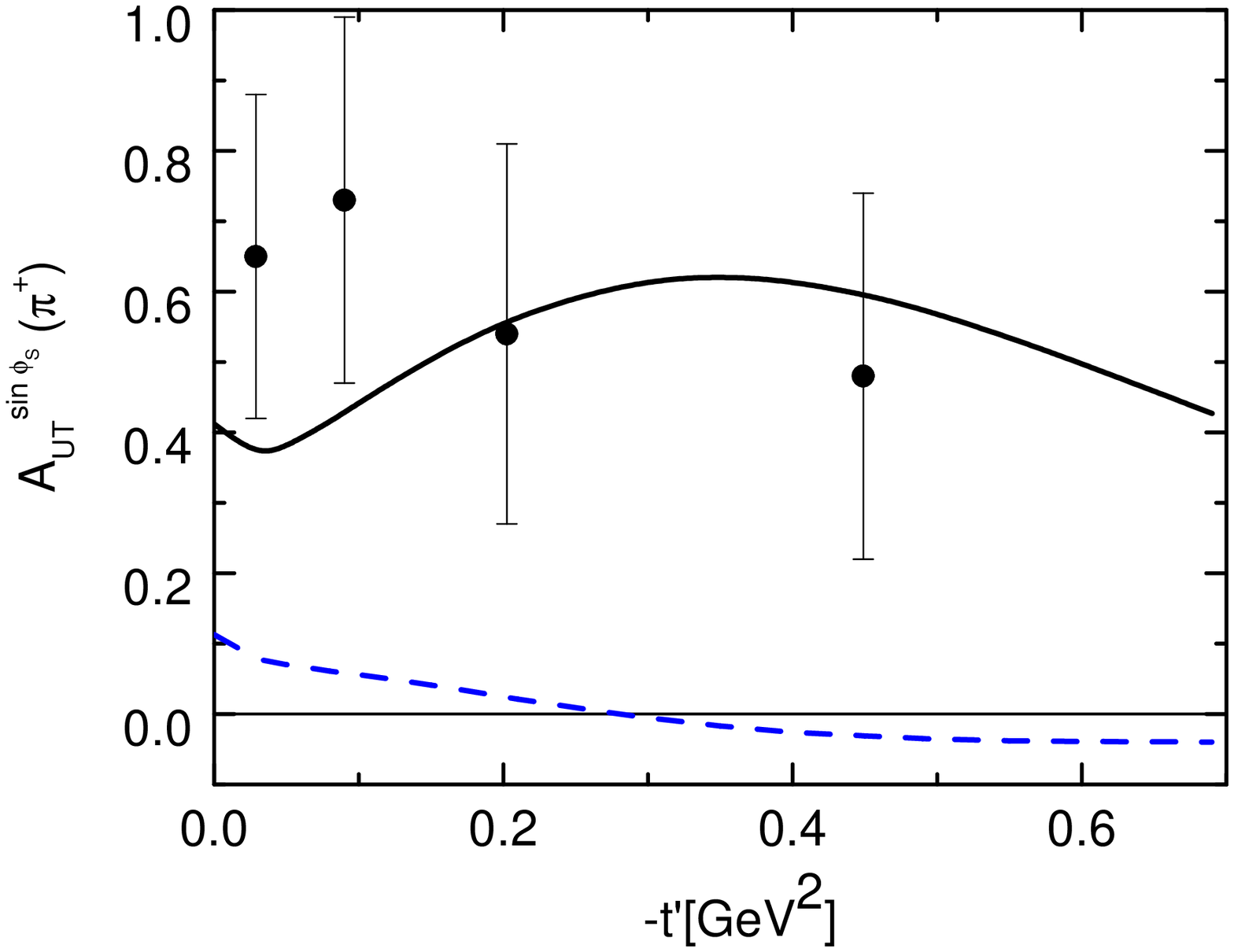,bb=23 350 532 743,width=0.46\textwidth,clip=true}
\hspace*{0.03\textwidth}
\psfig{file=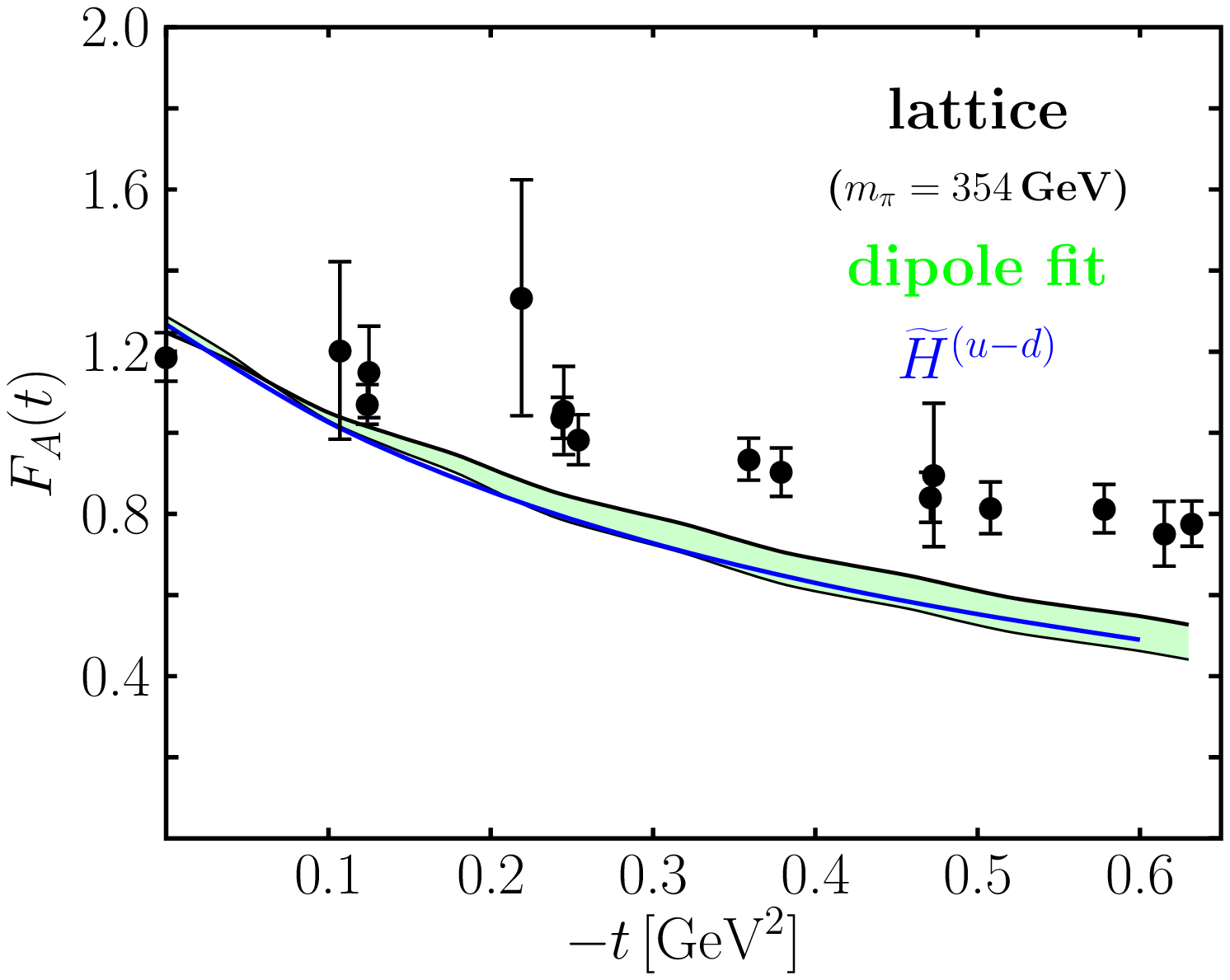,bb=158 372 590 717,width=0.45\textwidth,clip=true}
\caption{\label{fig:2} The $\sin{\phi_s}$ moment for a
   transversely polarized target at $Q^2\simeq 2.45\,{\rm GeV}^2$ and
    $W=3.99\,{\rm GeV}$ for $\pi^+$ production. The predictions from the handbag
    approach~\protect\cite{GK5} are shown as a solid line. The 
    dashed line is obtained under neglect of the twist-3 contribution. Data are 
    taken from Ref.\ \protect\refcite{Hristova}.}
\caption{\label{fig:3} The axial-vector form factor. The green band represents
  the dipole fit to the experimental data~\cite{kitagaki}, the solid circles the
  lattice results~\cite{haegler} for $m_\pi=354\,{\rm MeV}$. The thick (blue)
  curve is the form factor evaluated from $\widetilde{H}$ with the profile
  function of Ref.\ \refcite{DFJK4}.} 
\end{figure}

The electroproduction cross sections measured with a transversely or
longitudinally polarized target consist of many terms, each can be projected
out by $\sin{\varphi}$ or $\cos{\varphi}$ moments where $\varphi$ is a
specific linear combination of $\phi$, the azimuthal angle between the lepton 
and the hadron plane and $\phi_s$, the orientation of the target spin vector.  
A number of these moments have been measured recently~\cite{Hristova,hermes02}. 
A particularly striking result is the $\sin{\phi_S}$ moment. The data on it,
displayed in Fig.\ \ref{fig:2}, exhibit a mild $t$-dependence and do not show 
any indication for a turnover towards zero for $t^\prime\to 0$. This behavior 
of $A_{UT}^{\sin{\phi_s}}$ at small $-t^\prime$ can only be produced by an 
interference term between the two helicity non-flip  amplitudes ${\cal M}_{0+,0+}$ 
and ${\cal M}_{0-,++}$ which are not forced to vanish in the forward direction 
by angular momentum conservation. The amplitude ${\cal M}_{0-,++}$ has to be 
sizeable because of the large size of the $\sin{\phi_s}$ moment. We therefore
have to conclude that there are strong contributions  from $\gamma^*_T \to\pi$ 
transitions at moderately large values of $Q^2$.
\begin{table*}[t]
\renewcommand{\arraystretch}{1.4} 
\tbl{Features of the asymmetries for a transversally and longitudinally
  polarized target. The photon polarization is denoted by L (longitudinal) and
  T (transversal). Asymmetries under control of TT interference terms are not
  shown in the table; they are very small. 1: There is a second contribution
  for which the helicities of the outgoing proton are interchanged.}
{\begin{tabular}{|c|| c | c | c | c |}
\hline     
 observable  & dominant &   amplitudes  & low $t^\prime$ \\
        & interf. term  &   &  behavior \\[0.2em]   
\hline
$A_{UT}^{\sin(\phi-\phi_s)}$ &  LL  & ${\rm Im}\big[{\cal M}^*_{0-,0+}
                       {\cal M}_{0+,0+}\big]$ & $\propto \sqrt{-t^\prime}$   \\[0.2em]
$A_{UT}^{\sin(\phi_s)}$ & LT  &  ${\rm Im}\big[{\cal M}^*_{0-,++}{\cal M}_{0+,0+}\big]$  & const.  \\[0.2em]
$A_{UT}^{\sin(2\phi-\phi_s)}$ & LT & ${\rm Im}\big[{\cal M}^*_{0-,-+}
                             {\cal M}_{0+,0+}\big]^{\;1)}$ &  $\propto t^\prime$\\[0.2em]
\hline
$A_{UL}^{\sin(\phi)}$   & LT & ${\rm Im}\big[{\cal M}^*_{0-,++} {\cal M}_{0-,0+}\big]$ & 
$\propto \sqrt{-t^\prime}$   \\[0.2em]
$A_{LU}^{\sin(\phi)}$   & LT & ${\rm Im}\big[{\cal M}^*_{0-,++} {\cal M}_{0-,0+}\big]$ & 
$\propto \sqrt{-t^\prime}$   \\[0.2em]
$A_{LL}^{\cos(\phi)}$   & LT & ${\rm Re}\big[{\cal M}^*_{0-,++} {\cal M}_{0-,0+}\big]$ & 
$\propto \sqrt{-t^\prime}$   \\[0.2em]
\hline
\end{tabular}}
\label{tab:1}
\renewcommand{\arraystretch}{1.0}   
\end{table*}

How can this amplitude be modeled in the frame work of the handbag approach? 
From Fig.\ \ref{fig:1} where the helicity configuration of the amplitude 
${\cal M}_{0-,++}$ is shown, it is clear that contributions from the
usual helicity non-flip GPDs, $\widetilde{H}$ and $\widetilde{E}$, to this 
amplitude do not have the properties required by the data on the
$\sin{\phi_s}$ moment. For these GPDs the emitted and re-absorbed partons from 
the nucleon have the same helicity. Consequently, there are net helicity flips 
of one unit at both the parton-nucleon vertex and the subprocess. Angular 
momentum conservation therefore forces both parts to vanish as
$\sqrt{-t^\prime}$. Thus, a contribution from the ordinary GPDs to 
${\cal M}_{0-,++}$ vanishes $\propto t^\prime$. There is a second set of
leading-twist GPDs, the helicity-flip or transversity ones $H_T, E_T, \ldots$
~\cite{diehl01,hoodbhoy} for which the emitted and re-absorbed partons have
opposite helicities. As an inspection of Fig.\ \ref{fig:1} reveals the 
parton-nucleon vertex as well as the subprocess amplitude ${\cal H}_{0-,++}$ 
are now of helicity non-flip nature and are therefore not forced to vanish in 
the forward direction. The prize to pay is that quark and anti-quark forming
the pion have the same helicity. Therefore, the twist-3 pion wave function is 
needed instead of the familiar twist-2 one. The dynamical mechanism building 
up the amplitude ${\cal M}_{0-,++}$ is so of twist-3 accuracy. It has been 
first proposed in Ref.\ \refcite{passek} for wide-angle photo- and 
electroproduction of mesons where $-t$ is considered to be  the large 
scale~\cite{huang}.

Allowing only for $H_T$ as the only transversity GPD in an admittedly rough
approximation the twist-3 mechanism only contributes to the amplitude
\begin{equation}
{\cal M}_{0-,++}^{\rm twist-3} = e_0 \int_{-1}^1 dx H^{(3)}_T(x,\xi,t) {\cal H}_{0-,++}
\label{twist-3}
\end{equation}
For the calculation of the subprocess amplitude ${\cal H}_{0-,++}$ the twist-3
pion wave function is taken from Ref.\ \refcite{braun90} with the three-particle 
Fock component neglected~\cite{GK5}. This wave function contains a pseudo-scalar 
and a tensor component. The latter one provides a contribution to 
${\cal M}_{0-,++}$ which is proportional to $t^\prime/Q^2$ and, hence, neglected. 
The contribution from the pseudo-scalar component to ${\cal M}_{0-,++}$ has the 
required properties. It is proportional to the parameter $\mu_\pi=m^2_\pi/(m_u+m_d)$ 
which appears as a consequence of the divergency of the axial-vector current. 
Since $m_u$ and $m_d$ are current quark masses $\mu_\pi$ is large, actually 
$\simeq 2\,{\rm GeV}$ at the scale of $2\,{\rm GeV}$. Thus, although 
parametrically suppressed by $\mu_\pi/Q$ as compared to the longitudinal
amplitudes, the twist-3 effect is sizeable for $Q$ of the order of a few GeV.
\section{The GPDs at small skewness}
For $\pi^+$ electroproduction the GPDs, namely $\widetilde{H}$, the non-pole
part of $\widetilde{E}$ and the most important one of the
transversity GPDs, $H_T$, contribute in the isovector combination
\begin{equation}
F_i^{(3)} = F_i^u - F_i^d\,.
\end{equation}
The GPDs are constructed with the help of double distributions 
ansatz~\cite{rad98} consisting of the product of the zero-skewness GPDs and an
appropriate weight function, actually parameterized as a power of the valence
Fock state meson distribution amplitude. This weight function generates the 
skewness dependence of the GPD. It is important to note that other methods to
generate the skewness dependence, namely the Shuvaev transform~\cite{martin}
or the dual parameterization~\cite{semenov} lead to very similar results for 
the GPDs at small skewness. The zero-skewness GPDs in the double distribution
ansatz are assumed to be given by products of their respective forward limits
and Regge-like $t$ dependences, $\exp{[f_i(x,t) t]}$, with profile functions 
that read
\begin{equation}
f_i(x,t) = b_i-\alpha_i^\prime \ln{x}
\label{profile1}
\end{equation}
where $\alpha_i^\prime$ is the slope of an appropriate Regge trajectory (pole
or cut). These profile functions can be regarded as small-$x$ approximations 
of more complicated versions used for the determination of the zero-skewness
GPDs from the nucleon form factors~\cite{DFJK4}   
\begin{equation}
f_i(x,t) = b_i(1-x)^3 - \alpha_i^\prime(1-x)^3 \ln{x} +A_i x(1-x)^2
\label{profile2}
\end{equation}
which hold at all $x$. There is a strong correlation between $t$ and $x$ in 
this ansatz: the behavior of moments or convolutions of a GPD at small (large)
$-t$ is determined by the small (large) $x$ behavior of this GPD. It is to be 
stressed that the analysis performed in Ref.\ \refcite{DFJK4} as well as
recent results from lattice QCD~\cite{haegler} clearly rule out a factorization 
of the zero-skewness GPDs in $x$ and $t$.

The forward limit of $\widetilde{H}$ is given by the polarized parton 
distributions $\Delta q(x)$,  that of $H_T$ by the transversity distribution
$\delta(x)$ for which the results of an analysis of the asymmetries in 
semi-inclusive electroproduction have been taken~\cite{anselmino}. Finally,  
the forward limit of the non-pole part of $\widetilde{E}$ is parameterized as
\begin{equation}
\tilde{e}^{(3)}(x)= \widetilde{E}^{(3)}_{\rm n.p.}(x,\xi=t=0) = 
                        \widetilde{N}_{\tilde{e}}^{(3)}  x^{-0.48} (1-x)^5\,,
\label{Enp}
\end{equation}
in analogy to the PDFs. The normalization $\widetilde{N}_{\tilde{e}}^{(3)}$
is fitted to experiment. The full set of parameters used in the analysis of 
$\pi^+$ electroproduction can be found in Ref.\ \refcite{GK5}.

The full GPD $\widetilde{E}^{(3)}$ is the sum of the pole and non-pole 
contribution where the first one reads
\begin{equation}
\widetilde{E}^u_{\rm pole} =- \widetilde{E}^d_{\rm pole} = \Theta(|x|\leq \xi) 
             \frac{F_P^{\rm pole}(t)}{4\xi} \Phi_\pi((x+\xi)/(2\xi))\,.
\label{e-tilde-pole}
\end{equation}
Here, $\Phi_\pi$ is the pion's distribution amplitude and $F_P$ is the 
pseudo-scalar form factor of the nucleon being related to
$\widetilde{E}^{(3)}$ by the sum rule
\begin{equation}
\int^1_{-1} dx \widetilde{E}^{(3)}(x,\xi,t) = F_P(t)\,.
\end{equation}
The evaluation of the pion electroproduction amplitude from the graph shown on
the left hand side of Fig.\ \ref{fig:1} just using $\widetilde{E}_{\rm pole}$ 
leads to the pion-pole contribution as given in (\ref{eq:L-amplitudes}) but
with only the perturbative contribution to the pion's electromagnetic form 
factor occurring in the residue (\ref{residue}). Other graphs have to be 
considered in addition  for the pion pole, e.g.\ the Feyman mechanism. In
order to avoid this complication the pion-pole contribution is simply worked
out from the graph shown on the right hand side of Fig.\ \ref{fig:1}.
 
In summary: the GPDs used in Ref.\ \refcite{GK5} are valid at small skewness 
($\xi \,{\raisebox{-4pt}{$\,\stackrel{\textstyle <}{\sim}\,$}}\, 0.1\;$)
and are probed by experiment for 
$x \,{\raisebox{-4pt}{$\,\stackrel{\textstyle <}{\sim}\,$}}\, 0.6\;$.
Due to the double distribution ansatz they satisfy polynomiality and the
reduction formulas. It has also been checked numerically that the lowest 
moments of the GPDs $\widetilde{H}$ and $\widetilde{E}$ are in agreement with
the data on the axial-vector~\cite{kitagaki} and pseudo-scalar~\cite{choi}
form factors of the nucleon (see Fig.\ \ref{fig:3}) and respect various  
positivity bounds~\cite{poby,diehl-haegler}. Comparison with recent lattice 
QCD studies~\cite{haegler, goeckeler} reveals that there is good agreement 
with the relative strength of moments and their relative $t$  dependences. 
At small $t$ even the absolute values of the moments agree quite well but 
the $t$ dependence of the moments obtained from  lattice QCD are usually 
flatter than those from the GPDs and the form factor data. An exception is the 
lowest moment of $H_T$ for $u$ quarks for which we have a value that is 
about $25\%$ smaller than the lattice result. Similar observation can be made
for the GPDs $H$ and $E$ which have been constructed analogously and probed 
in vector meson electroproduction~\cite{GK1}. As an example the axial-vector 
form factor obtained from $\widetilde{H}$ as used in Ref. \refcite{GK5} is 
compared with experiment and with the lattice QCD results in Fig.\ \ref{fig:3}. 
 
\section{Results}
\begin{figure}[t]
  \begin{center}
  \psfig{file=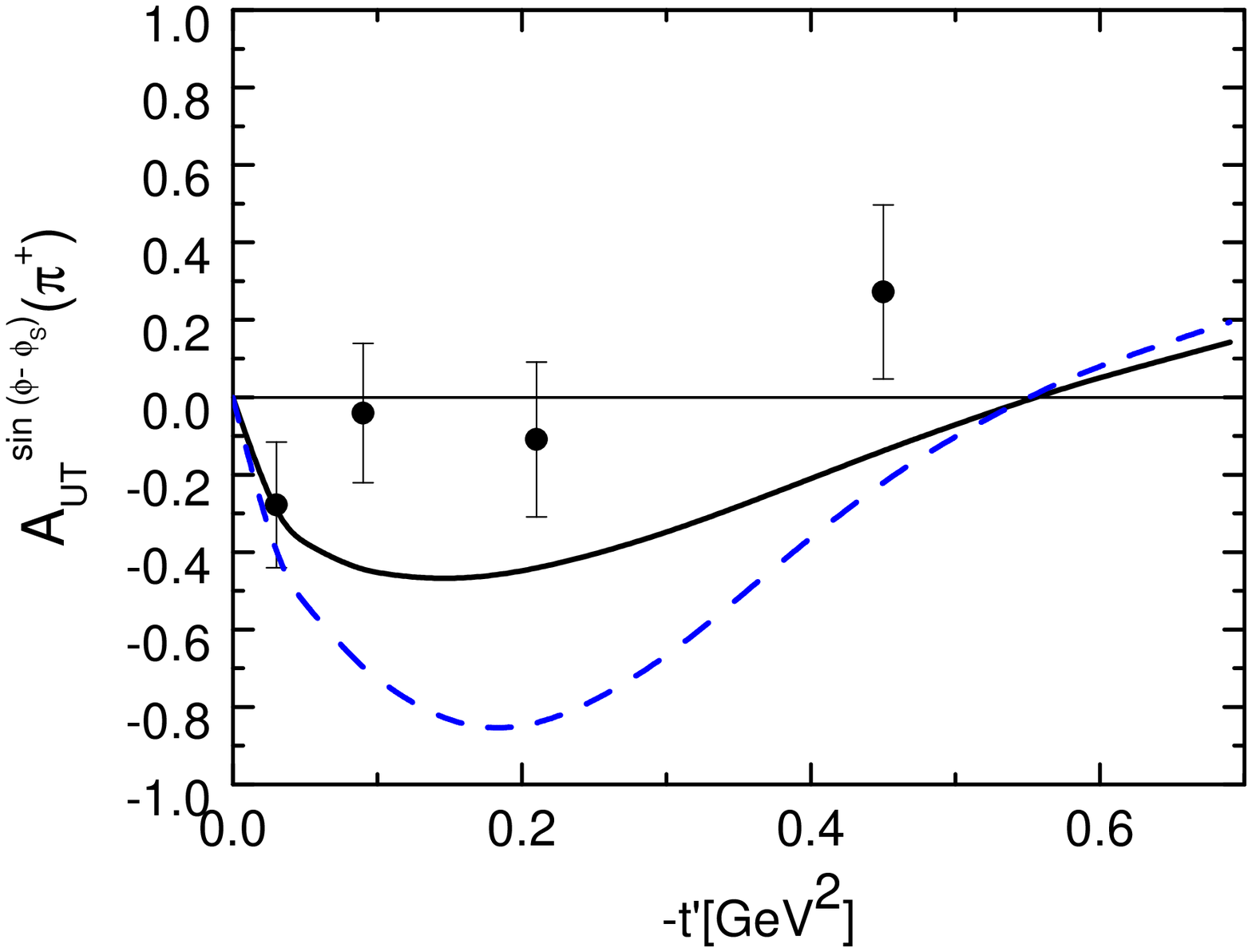,bb=16 349 533 743,width=0.44\textwidth,clip=true}
  \psfig{file=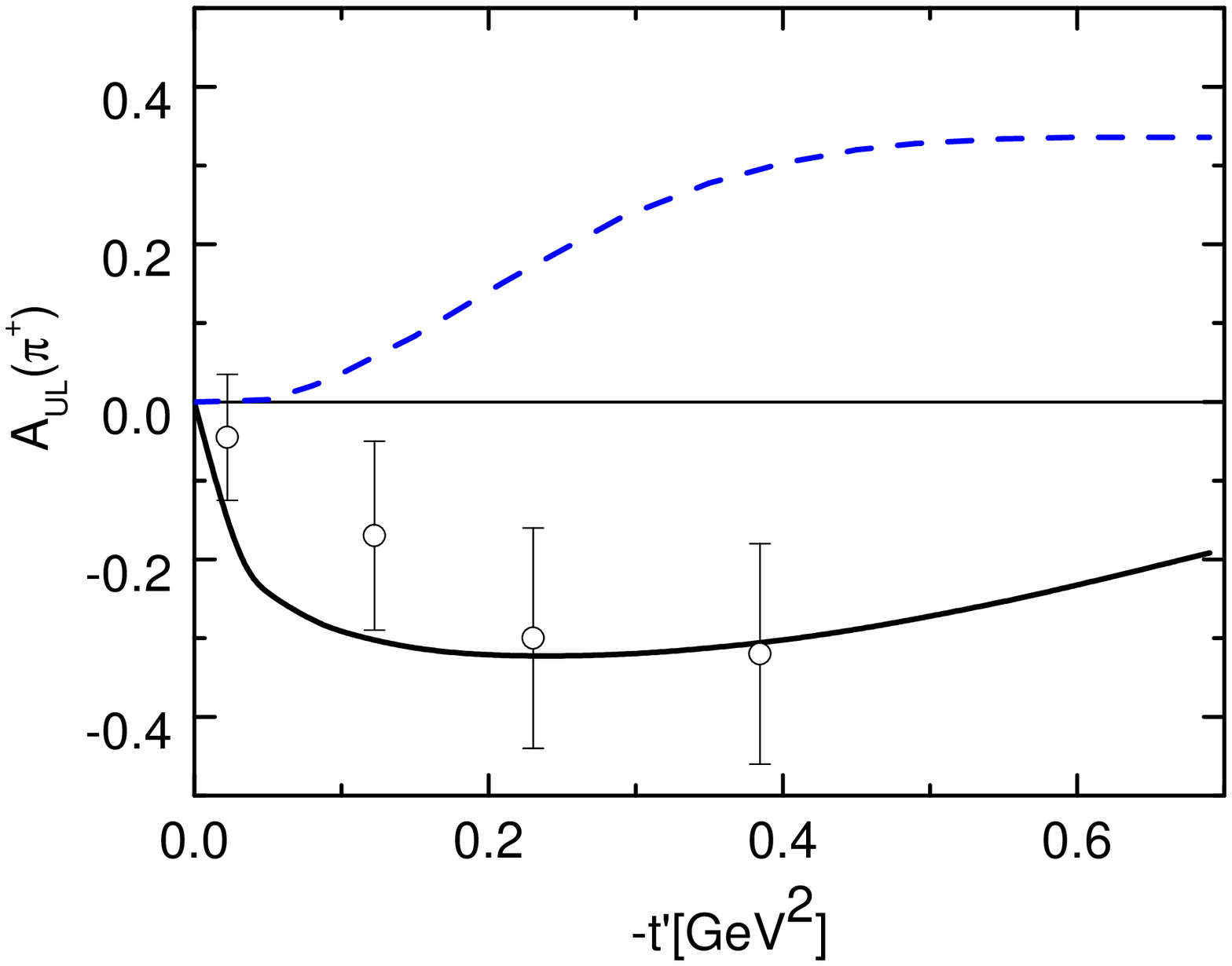,bb=30 346 533 746,width=0.42\textwidth,clip=true}
  \caption{ \label{fig:5} Left: Predictions for the  $\sin{(\phi-\phi_s)}$
    moment at $Q^2=2.45\,{\rm GeV}^2$  and $W=3.99\,{\rm GeV}$ shown as solid
    lines~\protect\cite{GK5}. The dashed line represents the longitudinal 
    contribution to the $\sin{(\phi-\phi_s)}$ moment. Data are taken 
    from \protect\cite{Hristova}.}
  \caption{\label{fig:6} Right:  The asymmetry for a longitudinally polarized 
  target at $Q^2\simeq 2.4\,{\rm GeV}^2$ and $W\simeq 4.1\,{\rm GeV}$. The
  dashed line is obtained disregarding the twist-3 contribution. Data are
  taken from \protect\cite{hermes02}.}
  \end{center}
  \end{figure}

It is shown in Ref.\ \refcite{GK5} that with the described GPDs, the $\pi^+$
cross sections as measured by HERMES~\cite{HERMES07} are nicely fitted  as
well as the transverse target asymmetries~\cite{Hristova}. This can be seen
for instance from Fig.\ \ref{fig:1} where $A_{UT}^{\sin{\phi_s}}$ is displayed. 
Also the $\sin(\phi-\phi_s)$ moment which is dominantly fed by an interference 
term of the two amplitudes for longitudinally polarized photons, is fairly
well described as is obvious from Fig.\ \ref{fig:5}. Very interesting is also 
the asymmetry for a longitudinally polarized target. It is dominated by an 
interference term between ${\cal M}_{0-,++}$ which comprises the twist-3 effect, 
and the nucleon helicity-flip amplitude for $\gamma^{\,*}_L\to \pi$ transition, 
${\cal M}_{0-,0+}$. Results for $A_{UL}^{\sin \phi}$ are displayed and compared 
to the data in Fig.\ \ref{fig:6}. In both the cases, $A_{UT}^{\sin{\phi_s}}$ and
$A_{UL}^{\sin \phi}$, the prominent role of the twist-3 mechanism is clearly
visible. Switching it off one obtains the dashed lines which are significantly 
at variance with experiment. In this case the transverse amplitudes are only 
fed by the pion-pole contribution. 

Although the main purpose of the work presented in Ref.\ \refcite{GK5} is
focused on the analysis of the HERMES data one may also be interested in 
comparing this approach to the Jefferson Lab data on the cross 
sections~\cite{horn06}. With the GPDs $\widetilde{H}, \widetilde{E}$ and $H_T$ in 
their present form the agreement with these data is poor. I remind the reader
that the approach advocated for in Refs.\ \refcite{GK5} and \refcite{GK1} is 
optimized for small skewness. At larger values of it the parameterizations 
of the GPDs are perhaps to simple and may require improvements as for instance
the replacement of the profile function (\ref{profile1}) by (\ref{profile2}). 
As mentioned above the GPDs are probed by the HERMES data only for $x$ less 
than about 0.6. One may therefore change the GPDs at large $x$ to some extent 
without changing much the results for cross sections and asymmetries in the 
kinematical region of small skewness. For Jefferson Lab kinematics, on the 
other hand, such changes of the GPDs may matter. Finally one should be 
aware that at larger values of skewness the other transversity GPDs may not 
be negligible. In a recent lattice study~\cite{haegler06} the moments of the 
combination $2\widetilde{H}_T+E_T$ have been found to be rather large in 
comparison to those of $H_T$. Including this combination of GPDs into the 
analysis of pion electroproduction one would have
\begin{equation}
{\cal M}_{0+,\mu +}^{\rm twist-3} = -e_0 \frac{\sqrt{-t^\prime}}{4m}
        \int_{-1}^1 dx \big[2\widetilde{H}^{(3)}_T+E_T^{(3)}\big] {\cal H}_{0-,++}
\label{twist-3-2}
\end{equation}
in addition to (\ref{twist-3}). Here, $\mu$ ($\pm 1$) labels the photon
helicity. The amplitude (\ref{twist-3-2}) holds up to corrections of order
$\xi$.

\section{Summary and outlook}
In summary, there is strong evidence for transversity in hard exclusive 
electroproduction of pions. A most striking effect is seen in the target
asymmetry $A_{UT}^{\sin \phi_s}$. The interpretation of this effect requires a 
large helicity non-flip amplitude ${\cal M}_{0-,++}$. Within the handbag 
approach this amplitude is generated by the helicity-flip or transversity 
GPDs in combination with a twist-3 pion wave function. This 
explanation establishes an interesting connection to transversity parton 
distributions measured in inclusive processes. Further studies of 
transversity in exclusive reactions are certainly demanded. Good data on 
$\pi^0$ electroproduction would also be welcome. They would not only allow 
for further tests of the twist-3 mechanism but also give the opportunity to 
verify the model GPDs $\widetilde{H}$ and $\widetilde{E}$ as used in Ref.\ 
\refcite{GK5}. An intriguing issue is whether or not the handbag approach 
in its present form for pion electroproduction works for the kinematics 
presently accessible at Jlab. It is known that it cannot accommodate the CLAS 
data on $\rho^0$, $\rho^+$ and $\omega$ production~\cite{clas}. Other
applications and tests of the handbag approach including the twist-3 mechanism
are pion electroproduction measured at the upgraded Jlab accelerator or by 
the COMPASS collaboration and the measurement of the time-like process 
$\pi^- p \to \mu^+\mu^-n$~\cite{pire}. The extension of this approach to 
electroproduction of other pseudoscalar mesons, in particular the $\eta$ and
$\eta^\prime$, is also of interest. In principle this would give access to the
GPDs for strange quarks. As has been shown in Ref.\ \refcite{passek03} there is
no complication in the analysis of the electroproduction data due to the 
two-gluon Fock components of the $\eta$ and $\eta^\prime$ since they are
suppressed by $t^\prime/Q^2$. 
 
\section{Acknowledgments}
The author wishes to thank Anatoly Radyushkin and Paul Stoler for inviting him
to present this talk at the workshop on Exclusive Reactions at High Momentum
Transfer. This work is supported in part by the BMBF under contract 06RY258.


\end{document}